\documentclass[10pt]{iopart}
\usepackage{graphicx}
\usepackage{dcolumn}
\usepackage{bm}
\usepackage{amssymb}
\usepackage{pifont}
\usepackage{times}
\usepackage{color}

\def \w {\omega}
\def \W {\Omega}
\def \e {\varepsilon}
\def \vp {\varphi}

\begin{document}

\title{A minimal model of partial synchrony}

\author{Pau Clusella}
\address{Institute for Complex Systems and Mathematical Biology, SUPA, 
University of Aberdeen, Aberdeen, UK}
\address{Dipartimento di Fisica, Universit\`a di Firenze, Italy}

\author{Antonio Politi}
\address{Institute for Complex Systems and Mathematical Biology, SUPA, 
University of Aberdeen, Aberdeen, UK}

\author{Michael Rosenblum}
\address{Department of Physics and Astronomy, University of Potsdam,
Karl-Libknecht-Strasse 24/25, 14476 Potsdam, Germany}

\date{\today}

\begin{abstract}
We show that self-consistent partial synchrony in globally coupled oscillatory
ensembles is a general phenomenon. We analyze in detail appearance and stability 
properties of this state in possibly the simplest setup of a
biharmonic Kuramoto-Daido phase model as well as demonstrate the effect 
in limit-cycle relaxational Rayleigh oscillators.
Such a regime extends the notion of splay state from a uniform to distribution
of phases to an oscillating one. Suitable collective observables such as the Kuramoto
order parameter allow detecting the presence of a inhomogeneous distribution.
The characteristic and most peculiar property of partial synchrony is the difference 
between the frequencies of single units and that of the macroscopic field.
\end{abstract}

\pacs{05.45.Jn, 05.45.-a}

\maketitle

\section{Introduction} 

Many physical systems can be represented as networks of oscillators, 
different examples ranging from the mammalian brain~\cite{Buzsaki-06},
to power grids~\cite{Fang-11}, out-of-equilibrium chemical reactions~\cite{Craciun-08},
spin-torque nanoscale 
oscillators~\cite{PhysRevB.73.060409,PhysRevB.82.140407,Pikovsky-13}, 
gene-controlled clocks in bacteria \cite{Prindle_etal-12}, and so on. 
A large number of books, chapters, and reviews devoted to the topic
testify to the importance of this 
subject~\cite{Winfree-80,Kuramoto-84,Strogatz-00,Pikovsky-Rosenblum-Kurths-01,%
Golomb-Hansel-Mato-01,Manrubia-Mikhailov-Zanette-04,Acebron-etal-05,Arenas200893,%
Breakspear-Heitmann-Daffertshofer-10,Pikovsky-Rosenblum-15}.

A general theory of oscillatory ensembles has not yet been developed. 
Indeed, such a theory requires taking into account many different features, 
such as the structure of the single units and their heterogeneity, 
as well as the topology and properties of the connections.
Even in the simple context of globally coupled \textit{identical} phase oscillators, 
it is not generally known what a kind of stationary regimes are to be expected.
Roughly speaking, they can be classified by referring to the distribution of
phases in the limit of a large number of oscillators (the so-called thermodynamic limit).
If the mutual interaction leads to phase attraction (at least
below a certain distance), the distribution of phases converges to a set of 
Dirac $\delta$'s which correspond to different clusters.
Full synchrony is the extreme case, where all 
oscillators converge to the same trajectory, i.e. to a single cluster.
In the presence of a mutual repulsion, a smooth phase distribution is 
observed instead. 
The splay state (or, equivalently, the asynchronous regime) is a 
prototypical example, characterized by a flat distribution of the phases
and absence of a collective mode.
Finally, one can encounter chimeras, where a big cluster coexists
with a group of non-synchronized units \cite{Yeldesbay-Pikovsky-Rosenblum-14}. 
Interestingly, in this state, the frequency of the cluster elements differs 
from the frequencies of asynchronous ones.

Most of the efforts have been devoted to the study of 
clustered~\cite{Okuda-93}
and chimera states \cite{Schmidt_etal-14,Sethia-Sen-14}
and much less to the identification and analysis of regimes 
characterized by a smooth but non-uniform distribution of phases.
Such regimes, typically characterized by a periodic collective evolution, are
herein referred to as self-consistent partial synchrony (SCPS).
The simplest form of SCPS is a ``rigid" rotation of the distribution, i.e. a regime
where the instantaneous frequency of the oscillators coincides with that of the
collective mode.
Such a regime can emerge if the coupling strength vanishes for some finite value
of the order parameter~\cite{Filatrella-Pedersen-Wiesenfeld-07,Giannuzzi_et_al-07,%
Pikovsky-Rosenblum-09}. 
In a less trivial form of SCPS (of primary interest here) the (average) frequency of 
the single units and that of the mean field differ from each other.
Moreover, the two frequencies are generally mutually incommensurate, i.e. no 
locking phenomena are observed, when a control parameter is continuously varied.
Thus, the microscopic dynamics is quasiperiodic.
Examples of such dynamics are: 
integrate-and-fire (IF) neurons interacting through finite-width pulses 
(the so-called $\alpha$-functions)~\cite{vanVreeswijk-96};
nonlinearly coupled Stuart-Landau systems~\cite{Rosenblum-Pikovsky-15};
a Kuramoto-like model obtained via phase reduction from the above mentioned
Stuart-Landau ensemble.
A variant of quasiperiodic partially synchronous dynamics has 
been detected in models beyond phase approximation, i.e. in globally 
coupled Hindmarsh-Rose neurons and Stuart-Landau oscillators;  
here the macroscopic and microscopic frequencies are equal only on average, 
but the motion of oscillators is additionally modulated by a generally incommensurate 
frequency~\cite{Ehrich-Pikovsky-Rosenblum-13,Rosenblum-Pikovsky-15}.

In the weak-coupling limit oscillators are effectively
described by a Kuramoto-Daido phase model \cite{Daido-93,Daido-93a,Daido-96,Daido-96a}
with a suitable coupling function 
$G(\Delta \phi)$, where $\Delta \phi$ is the phase difference between any 
two interacting oscillators.
In the standard, widely used, Kuramoto-Sakaguchi 
model~\cite{Kuramoto-75,Kuramoto-84,Sakaguchi-Kuramoto-86} 
the coupling function $G$ is assumed to be perfectly sinusoidal. In the last
years it has become increasingly clear that this is quite a special case: 
e.g., multiple clusters in this setup 
are not possible~\cite{Engelbrecht-Mirollo-14,Pikovsky-Rosenblum-15}. 
A much richer dynamics, including formation of clusters~\cite{Okuda-93} and 
of heteroclinic cycles~\cite{Hansel-Mato-Meunier-93},  
is observed as soon as just one additional harmonic is added to $G$.

In this paper we further illustrate the richness of the 
Kuramoto-Daido model, by showing that SCPS spontaneously
emerges in a minimal extension of the Kuramoto setup, where $G$ is composed of 
just two harmonics.
To further explore the ubiquity of SCPS,
we study its emergence in an ensemble of linearly coupled Rayleigh oscillators. 
They are two-dimensional limit-cycle oscillators; performing numerically 
the reduction to a Kuramoto-Daido phase model, we reconstruct the coupling function  
which turns out to contain a few harmonics. The Kuramoto-Daido setup is shown to
reproduce the dynamics of the original system.

The simplicity of the biharmonic model allows for a detailed analysis of SCPS, 
which can be seen as a stationary solution of a continuity equation in a suitably 
rotating frame. 
Accordingly, the phase-distribution can be accurately determined from 
the emergence of SCPS 
out of the splay state -- through a Hopf bifurcation -- 
to its collapse onto full synchrony 
as in~\cite{Rosenblum-Pikovsky-07,Pikovsky-Rosenblum-09}. 
The additional stability analysis confirms the numerical evidence of the onsed of an instability and 
allows identifying the unstable direction.

As SCPS in the biharmonic model coexists with two-cluster states, we 
revisit their stability properties, to understand under which conditions 
trajectories converge towards an heteroclinic cycle 
(HC)~\cite{Hansel-Mato-Meunier-93,Kori-Kuramoto-01}.
We find that more hamornics are needed to 
ensure that two-cluster states and the corresponding HCs are both unstable.
Moreover, we find that heteroclinic cycles can be viewed at as a kind
of quasiperiodic partial synchrony.

The paper is organized as follows.  In Sec. II the model is briefly
introduced and the conditions for the stability of the fully synchronous and
asynchronous regimes are recalled. The corresponding phase-diagram is thereby
presented for a fixed amplitude of the second harmonic.
In Sec. III, we analyze the occurrence of SCPS, show how it can be treated and 
finally develop the
formalism needed to perform the stability analysis. 
Sec. IV is devoted to a discussion of two-cluster states and HC
analyzed in \cite{Hansel-Mato-Meunier-93,Kori-Kuramoto-01}. 
Here, after briefly recalling some known properties, we present a general 
analysis of the stability properties of two-cluster states,
in the perspective of shedding light
on the general conditions under which such states can be effectively unstable
(this is, for instance the case of the LIF model proposed in~\cite{vanVreeswijk-96}).
The theoretical predictions are then extensively tested in Sec. V.
There we find that the mean-field frequency is not necessarily smaller than the
frequency of the splay state as in the LIF model.
Additionally we discover that SCPS can lose stability (through a Hopf bifurcation)
and thereby lead to a collapse onto HCs. Actually, in order to avoid spurious effects 
due to finite
computer accuracy, a minimal heterogeneity is added which makes the oscillators slightly
different from one another. As a result, we find a bistable regime, where SCPS coexists with
stable HC.
In Sec. VI, we discuss SCPS in two other setups that can be effectively described 
by a suitable Kuramoto-Daido model: a set of LIF neurons, and an ensemble of 
Rayleigh oscillators.
A Kuramoto-Daido description of the former model was already presented in 
Ref.~\cite{Politi-Rosenblum-15} where the validity of the approximation was investigated.
Here we analyze two-cluster states verifying their instability.
As for the Rayleigh oscillators, we reduce their description to 
a Kuramoto-Daido phase model and
show that, at variance with the other setups formerly considered,  here instability 
of the splay state is due to harmonics higher than the second one.
The main results and the still open problems are summarized in the last section.

\section{The model}
We hereby consider the Kuramoto-Daido type model of identical all-to-all coupled 
phase oscillators. Performing a transformation to the co-rotating coordinate frame 
we set the frequency to zero, so that the model reads
\begin{equation}
\label{eq2}
  \dot{\phi}_i= \frac{1}{N} \sum_j G(\phi_j -\phi_i) \, ,
\end{equation}
where the coupling constant has been eliminated by performing
a suitable rescaling of time.
In most of the paper we consider the coupling function 
\begin{equation}
G(\phi) = \sin (\phi+\gamma_1) + a \sin(2\phi+\gamma_2) \; .
\label{eq1}
\end{equation}
A standard way to classify the configurations of an ensemble of oscillators 
is via the set of complex order parameters
\begin{equation}
Z_m=R_me^{i\beta_m}=\frac{1}{N}\sum_j \mathrm{e}^{im\phi_j} \;,
\label{genpar}
\end{equation}
where $R_1$ is the famous Kuramoto order parameter~\cite{Kuramoto-75,Kuramoto-84}, 
which is equal to $1$ in 
the case of full synchrony and is equal to $0$ in splay states. 

By making use of this definition, Eq.~(\ref{eq2}) can be rewritten as
\begin{equation}
\dot \phi_k = R_1 \sin(\beta_1-\phi_k+\gamma_1)+aR_2\sin(\beta_2-2\phi_k+\gamma_2)\;.
\label{eq3}
\end{equation}
This model was already considered in \cite{Hansel-Mato-Meunier-93,Kori-Kuramoto-01} 
with an emphasis on analysis of clustered states (see the next section) and has 
recently attracted a growing
interest, especially in the presence of a distribution of oscillator
frequencies \cite{Komarov-Pikovsky-13a,Komarov-Pikovsky-14}.

For $a=0$ Eqs.~(\ref{eq1},\ref{eq2}) reduce to the famous Kuramoto-Sakaguchi model.
In this case it is known that the fully synchronous solution $\phi_k=\phi$ is stable if and
only if $|\gamma_1|<\pi/2$, while the stability condition of the splay state 
is exactly opposite: asynchrony is stable for $\pi/2<|\gamma_1|<3\pi/2$ and unstable otherwise. 
A partially synchronous solution can arise only at the border 
between stability and instability of the two regimes.

This pointwise region can be made structurally stable 
by assuming that $\gamma_1$ is a 
function of the order parameter $R_1$ and of the coupling 
strength $\e$~\cite{Rosenblum-Pikovsky-07,Pikovsky-Rosenblum-09}
(such phase model can be obtained in the process of phase reduction of a system
of nonlinearly coupled Stuart-Landau oscillators~\cite{Rosenblum-Pikovsky-15}). 
The resulting regime was called self-organized quasiperiodic dynamics.

\begin{figure}
\begin{center}
\includegraphics[width=0.45\textwidth,clip=true]{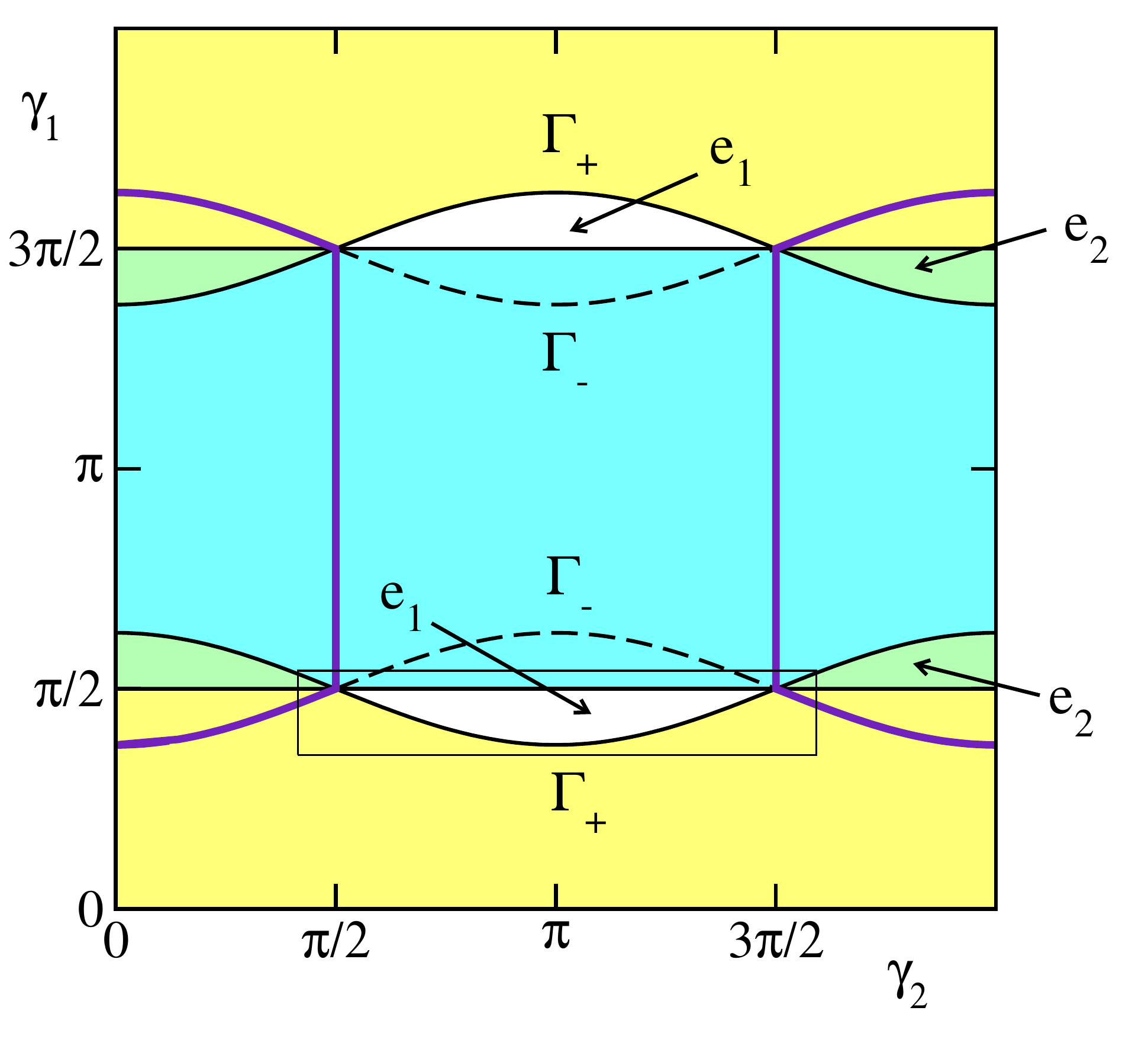}
\caption{Stability diagram of the biharmonic model (\ref{eq3})
for $a=0.2$. The splay state is stable
in the rectangular region $\pi/2<\gamma_1<3\pi/2$ (cyan and green areas);
the fully synchronous solution is stable above the upper and below the lower 
solid curves (yellow and green areas); two-cluster (anti-phase) states are 
stable inside the area delimited by the purple curve 
(i.e. below $\gamma_2=\pi/2$ and above $\gamma_2=3\pi/2$).
$\Gamma_+$ and $\Gamma_-$ identify two pairs of sinusoidal curves where the synchronous
state and the antiphase two-cluster states solution lose stability, respectively.
$e_1$ and $e_2$ denote two pairs of eyelets (bounded by $\Gamma_+$ and
$\Gamma_-$) where nontrivial dynamics between synchrony and asynchrony can be expected.
The box identifies the parameter region numerically investigated in this paper, see 
Fig.~\ref{fig:2}.
}\label{fig4}
\end{center}
\end{figure}

In this paper we show that adding a second harmonic is yet a simpler way to generate SCPS.
In the presence of a non-zero $a$, the stability of the fully synchronous state is determined
by the condition
\begin{equation}
G'(0) = -\cos\gamma_1 - 2a \cos\gamma_2 < 0\;.
\end{equation}
The marginal stability line is composed of two sinusoidal
curves $\Gamma_+$, shown in the stability diagram in Fig.~\ref{fig4}. 
Altogether, the synchronous state is stable in the two yellow and green regions
of the diagram.

On the other hand, the stability of the asynchronous state, which we analyze in the
thermodynamic limit $N\to\infty$, 
is still determined only by the amplitude of the first 
mode of the perturbation (see \cite{Politi-Rosenblum-15})
\begin{equation}
\delta_1 = \pi [\cos \gamma_1 + i \sin \gamma_1] \; ,
\label{eq:sakasplay}
\end{equation}
so that it depends only on $\gamma_1$. As a result, the splay state
is stable in a rectangular region where $\cos \gamma_1<0$ 
(see the cyan and green areas in Fig.~\ref{fig4}).
In the green areas, both the splay and the synchronous states  
are simultaneously stable, while in the white areas both of them are unstable. 

Altogether, the diagram in Fig.~\ref{fig4} has a reflectional symmetry with respect
to both the horizontal and vertical semi axes. Physically, there is only one symmetry:
if $\phi \to -\phi$, the same dynamics is found upon mapping $\gamma_1 \to 2\pi -\gamma_1$ 
and $\gamma_2 \to 2\pi -\gamma_2$. The additional symmetry is, therefore, only accidental.

\section{Partial synchronization}
\label{sec:PS}
In the thermodynamic limit one can investigate the various regimes by studying
the evolution of the probability density $P(\phi,t)$ of oscillators with 
phase $\phi$ at time $t$. It satisfies the continuity equation
\begin{equation}
\frac{\partial P}{\partial t}  =
- \frac{\partial }{\partial \phi} \left[ \left(\int d\psi G(\psi-\phi)P(\psi,t)\right) P(\phi,t) \right]\;.
\label{eq:partialKD}
\end{equation}
The splay state corresponds to a flat and constant density $P=1/(2\pi)$. As already illustrated in
Ref.~\cite{Politi-Rosenblum-15}, SCPS typically manifests itself as a rotating nonuniform density, which
emerges past a Hopf bifurcation. It is therefore convenient to perform a change of 
variables, introducing the angle $\theta = \phi - \Omega t$ and $Q(\theta,t) = P(\phi,t)$.
The corresponding evolution equation takes the form
\begin{equation}
\frac{\partial Q}{\partial t}  =
\frac{\partial }{\partial \theta} \left[\left(\Omega \!-\! 
\int d\psi G(\psi-\theta)Q(\psi,t)\right) Q(\theta,t) \right]\;.
\label{eq:partialKD2}
\end{equation}
For a suitably chosen $\Omega$ there exists a stationary time-independent solution 
$Q_0(\theta)$, which satisfies the equation 
\begin{equation}
\left[ \Omega- \int d\psi G(\psi-\theta)Q_0(\psi)\right] Q_0(\theta) = \eta \;,
\label{eq:partialfix}
\end{equation}
where $\eta$ is the probability flux. Notice that $2\pi \eta$ 
can be interpreted as the average microscopic frequency of the single oscillators 
in the moving frame.
Upon expanding $Q_0(\psi)$ in Fourier modes,
\begin{eqnarray}
Q_0(\psi) = \sum_m Z_m \mathrm{e}^{-i m\psi} \; ,
\end{eqnarray}
(the harmonics coincide with the generalized order parameters defined in
Eq.~(\ref{genpar}) for a finite-size ensemble of oscillators),
we can solve Eq.~(\ref{eq:partialfix}), obtaining
\begin{equation}
Q_0(\theta)\! =\! \frac{\eta}{\pi i\!\left[ 
Z_1 \mathrm{e}^{-i(\theta-\gamma_1)} \! + \!
aZ_2 \mathrm{e}^{-i(2\theta-\gamma_2)}\!-\!c.c.\right]\!+\!\W} \; .
\label{eq:partial_ana}
\end{equation}
Since the phase of the solution is arbitrary, we are free to fix it by imposing that $Z_1$ is real.
By considering that $\eta$ can be determined by imposing a normalization on $Q_0$, the above
equation contains four unknowns: $\W$, $Z_1$, and $Z_2$ (the last variable
is complex). They can be determined self-consistently by imposing
\begin{equation}
 Z_k =\frac{1}{2\pi} \int d\psi \mathrm{e}^{ik\psi} Q_0(\psi) 
\label{eq:cond}
\end{equation}
for $k=1,2$.
The solution can be found by searching for a fixed point in a four-dimensional space.

\subsection{Stability analysis}
Consider an infinitesimal perturbation $q(\theta,t)$ of $Q_0(\theta)$
and linearise Eq.~(\ref{eq:partialKD2}), making use of Eq.~(\ref{eq:partialfix}). One
obtains
\begin{equation}
\frac{\partial q(\theta,t)}{\partial t}  =
\frac{\partial }{\partial \theta} \left[\eta \frac{q(\theta,t)}{Q_0(\theta)}\! -\! Q_0(\theta)\!
\int d\psi G(\psi-\theta)q(\psi,t) \right] \;.
\label{eq:SCPSlinear1}
\end{equation}
By expanding the perturbation in Fourier series, 
\[
\hat q_k =\frac{1}{2\pi} \int d\psi q(\psi,t)\mathrm{e}^{ik\psi}
\]
one can rewrite the integral in the previous equation as
\[
\int\!\! d \psi G(\psi -\theta) q(\psi,t)\! =\! -\pi i \!
\left (\! \hat q_1 \mathrm{e}^{-(\theta-\gamma_1)i}\! + \!
\hat q_2 a \mathrm{e}^{-(2\theta-\gamma_2)i}\! -\!c.c. \!\right)\! \equiv\! B(\theta) 
\]
so that,
\begin{equation}
\frac{\partial q(\theta,t)}{\partial t}  =
\frac{\partial }{\partial \theta} \left[\eta \frac{q(\theta,t)}{Q_0(\theta)} 
- Q_0(\theta) B(\theta)
\right]\;.
\label{eq:SCPSlinear2}
\end{equation}
This equation can be expressed as a Fourier series.
With the help of Eq.~(\ref{eq:partial_ana}), it is found
\begin{eqnarray}
&&\hspace{-2.cm} \frac{ d \hat q_m}{dt} = m \left \{- i \W \hat q_m + 
\pi \mathrm{e}^{i\gamma_1} \left[ Z_1 \hat q_{m-1} + Z_{m-1} \hat q_{1} \right ] - 
\pi\mathrm{e}^{-i\gamma_1} \left[ Z_{-1} \hat q_{m+1} + Z_{m+1} \hat q_{-1} \right ] + \right .\nonumber \\
&&\hspace{-1.cm}\left . a\pi\mathrm{e}^{i\gamma_2} \left[ Z_2 \hat q_{m-2}+ Z_{m-2} \hat q_{2} \right] - 
 a\pi\mathrm{e}^{-i\gamma_2} \left[ Z_{-2} \hat q_{m+2} + Z_{m+2} \hat q_{-2}\right] \right \} \; .
\label{eq:lineargen} 
\end{eqnarray}
The $m$-th Fourier mode of the perturbation is coupled with the four nearest neighbour modes 
($m-2$, $m-1$, $m+1$, and $m+2$) as well as with the first two modes; the latter 
coupling is mediated by the amplitude of
higher components of the stationary solution $Q_0$. In other words, the corresponding matrix is sparse: it is
pentadiagonal with two full rows. As each derivative is multiplied by $m$, $\dot {\hat q}_0=0$. 
This is a straightforward
consequence of the conservation of the total probability; therefore $\hat q_0$ can be eliminated as it 
does not contribute to the eigenvalues.
By further looking at the evolution equation for $\hat q_2$, we see that it involves $\hat q_{-1}$, so that
the negative modes must be included as well. Since $\hat q_{-m} = \hat q_m^*$, it is convenient to separate
$\hat q_m$ into real and imaginary part ($\hat q_m = u_m + i v_m$), so that we can exploit the relationships
$u_{-m}=u_m$, $v_{-m} = - v_m$ and thereby get rid of the negative $m$ components. 
The relevant eigenvalues $\mu_R+i\mu_I$ of the resulting matrix can be then computed by considering a sufficiently
large number of Fourier modes.

\section{Clusters and heteroclinic cycles}
\label{sec:clusters}

Clustered states represent another class of stationary solutions. In the biharmonic model
such states have been already investigated in~\cite{Hansel-Mato-Meunier-93,Kori-Kuramoto-01}. 
Here below we summarize those results that are necessary to proceed ahead with our general
considerations.

A two-cluster state consists of two families of oscillators with phases $\alpha$ and $\psi$,
respectively. 
Both for the sake of simplicity and since they are the most widely observed,
here, we mostly focus on the symmetric case of equal-size clusters.
The phases of the two families follow the differential equations,
\begin{eqnarray}
\label{eq:cluster1}
  \dot{\alpha} = (G(0) + G(\psi-\alpha))/2 \quad , \quad
  \dot{\psi} = (G(\alpha-\psi) + G(0))/2 \; .
\end{eqnarray}
The separation $\delta = \alpha - \psi$ satisfies the equation
\begin{eqnarray}
\label{eq:cluster2}
  \dot{\delta} &=& (G(-\delta) - G(\delta))/2 = -G_A(\delta) \; ,
\end{eqnarray}
where $G_A$ is the anti-symmetric component of $G$.
Two-cluster solutions are identified by the zeros of $G_A$;
$\delta=0$ is always a solution which corresponds to a single cluster 
(vanishing distance between the two clusters).

In the biharmonic model, the symmetric and anti-symmetric component of the coupling function are
\begin{eqnarray}
G_A(\delta) &=& \sin \delta (\cos \gamma_1 + 2 a \cos \gamma_2 \cos \delta)\;,
\label{eq:twomo1} \\
G_S(\delta) &=& \sin \gamma_1 \cos \delta + a \sin \gamma_2 \cos 2\delta\;.
\label{eq:twomo2}
\end{eqnarray}
There are various solutions of the equation $G_A(\delta)=0$ besides $\delta_0=0$:
$\delta_0=\pi$ corresponds to an antiphase two-cluster state. Two further solutions can
be found by setting to zero the expression in parentheses in Eq.~(\ref{eq:twomo1});
however, these solutions represent only one physically meaningful state. Indeed, given a
two-cluster state characterized by a separation $\delta_0$, the same state can be
seen as characterized by a separation $2\pi-\delta_0$, if the two clusters are exchanged.
These  states exist only in the parameter region delimited by the curves where
\begin{eqnarray}
\cos \gamma_1 \pm 2a \cos \gamma_2 = 0 \, .
\label{eq:clust3}
\end{eqnarray}
This equation with a plus sign defines the curve $\Gamma_+$ which coincides
with the bifurcation line where the synchronous state loses stability, see 
Fig.~\ref{fig4}.
(In fact, $\cos \delta_0 = 1$ means $\delta_0=0$, i.e.
the two-cluster solution bifurcates from the fully synchronous one.) 
The minus sign, instead, corresponds to the curve $\Gamma_-$ where the two-cluster 
state becomes the antiphase one.
As a result, nontrivial 
clustered solutions with $\delta\ne \pi$ exist only in the regions delimited by 
$\Gamma_+$ and $\Gamma_-$ (see the two pairs of eyelets $e_1$ and $e_2$ 
in Fig.~\ref{fig4}).

The stability of a two-cluster state is determined by the value of 
the inter- and intra-cluster exponents.
The inter-cluster exponent $\lambda_I$ measures the stability against perturbation
of the phase-separation between the two clusters. From Eq.~(\ref{eq:cluster2}) it follows
\begin{eqnarray}
\label{eq:clust_stab0}
\lambda_I = -G'_A(\delta_0) \; .
\end{eqnarray}
In the biharmonic model,
$G'_A(\delta_0) = \cos \gamma_1 \cos \delta_0 + 2a \cos \gamma_2 \cos 2\delta_0$.
The intra-cluster exponents $\lambda_E^{\pm}$ measure the stability of the width of
each cluster. From the equations of motion it is readily found that
\begin{equation}
\label{eq:clust_stab}
\lambda_E^{\pm}  \! =\! -[G'_A(0)\! +\! G'_A(\delta_0)\! 
  \pm G'_S(\delta_0)]\ \;.
\end{equation}
The solution with the plus (minus) sign refers to the cluster that is lagging behind (leading)
by $\delta_0$.
The maximal eigenvalue is therefore,
\begin{eqnarray}
\label{eq:clust_stabi2}
  \lambda_M = -\left[G'_A(0) + G'_A(\delta_0) - |G'_S(\delta_0)|\right ]/2 \; .
\end{eqnarray}
The inter-cluster exponent of the antiphase solution,
$G'_A(\pi) = -\cos \gamma_1 + 2a \cos \gamma_2$, 
is negative in between the upper and lower branch of $\Gamma_-$ and vanishes
along $\Gamma_-$, confirming that the clustered solutions bifurcate out of the 
antiphase state.  This region is almost complementary to the stability area
of the synchronous state, since $G'(\pi)G'(0)<0$ in the region where other clusters 
do not exist.
The stability of the antiphase state to intra-cluster perturbations 
is controlled by (see Eq.~(\ref{eq:clust_stabi2})),
\begin{eqnarray}
\label{eq:clust_stabi3}
  \lambda_M = -2a \cos \gamma_2 \; ,
\end{eqnarray}
so that this state is stable within the region
delimited by the purple curve in Fig.~\ref{fig4}.

The nontrivial two-cluster state ($\delta_0\ne \pi$) is unstable against inter-cluster
perturbations within the eyelets $e_2$ (because of the one-dimensional nature of the equation for $\delta$, 
it has opposite stability with respect to that of the synchronous solution), while it is stable within $e_1$.
Its intra-cluster stability can be determined from Eq.~(\ref{eq:clust_stabi2}). By taking into account 
that $G_A(\delta_0)=0$, we find that
\begin{eqnarray}
2\lambda_M &=& -\cos \gamma_1(1- \cos \delta_0) + 
|\sin \delta_0(\sin \gamma_1 + 4a \sin \gamma_2 \cos \delta_0) | \; . 
\label{eq:clust_gen}
\end{eqnarray}
Therefore, such a two-cluster state is unstable for $\pi/2<\gamma_1<3\pi/2$. A detailed analysis for
$a=0.2$ reveals that the solution is unstable also within the eyelet $e_1$.

This is not yet the end of the story. The opposite sign of the two exponents $\lambda_E^{\pm}$
implies that, while the width of one cluster decreases, that of the other one diverges.
However, as already discussed in~\cite{Hansel-Mato-Meunier-93}, once the width of the ``exploding" 
cluster becomes of order 1, nonlinear effects (not captured by a linear stability analysis) 
induce a relative phase shift of the two clusters, so that the leading cluster becomes the lagging one:
this implies a stability ``exchange".
As a consequence, the long term behavior can be assessed after averaging over the alternating periods
of stability and instability. Symmetry reasons imply that the time duration of such two periods
are equal to one another, so that the average exponent is, in general,
\begin{eqnarray}
\label{eq:clust_staba}
  \lambda_a = -\left[G'_A(0) + G'_A(\delta_0) \right ]/2 \; .
\end{eqnarray}
In the biharmonic model
\begin{equation}
  \lambda_a = -\cos \gamma_1(1- \cos \delta_0)/2 \; .
\label{eq:intra_biha}
\end{equation}
In the region of interest, $\lambda_a<0$ if $\cos \gamma_1>0$, i.e the fluctuations of the cluster
widths on average decrease, without ever collapsing onto it.
This is nothing but an attracting heteroclinic cycle (HC).
Because of the finite computer accuracy, these oscillations necessarily collapse on
the otherwise unstable two-cluster state.

\section{Numerical simulations}

\subsection{Microscopic analysis}

We start by exploring the parameter region identified by the 
rectangle in Fig.~\ref{fig4}, 
which includes the area where highly symmetric synchronous and asynchronous states
and two-cluster states are all unstable. 
(Because of the above mentioned symmetry, the upper eyelet is 
characterized by an equivalent dynamics.)

\begin{figure}
\begin{center}
\includegraphics[width=0.6\textwidth,clip=true]{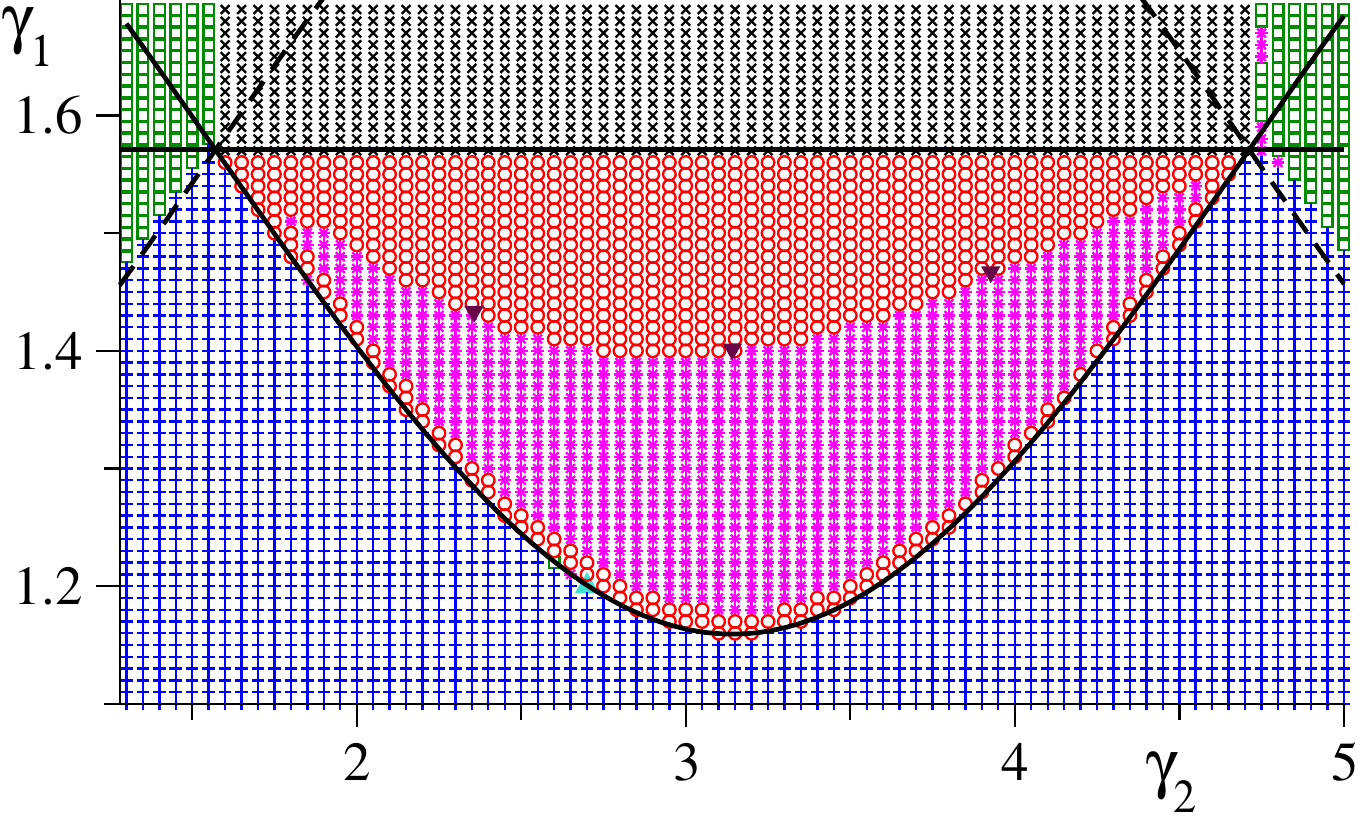}
\caption{Map of dynamical regimes of the system ~(\ref{eq2},\ref{eq1}),
obtained via a direct numerical simulation, starting from a slightly 
perturbed splay state.
The explored area corresponds to the rectangle in 
Fig.~\ref{fig4}. Notations are as follows.
Black crosses: asynchronous solutions, blue pluses: synchrony, 
green square: two-cluster states, red circles: SCPS, 
magenta stars: heteroclinic cycles. 
Cyan triangle at $\gamma_2=2.7$, $\gamma_1=1.2$ marks a three-cluster
state.
Black theoretical curves are the same as in Fig.~\ref{fig4}.
The map has been computed for 1000 oscillators, with the transient time 
$2\cdot 10^4$.  The three triangles correspond to points where SCPS is
marginally stable as obtained from Eq.~(\ref{eq:lineargen}).
} 
\label{fig:2}
\end{center}
\end{figure}

The outcome of a direct integration of Eqs.~(\ref{eq2},\ref{eq1}) 
(starting from an initial
condition close to the splay state) is summarized in Fig.~\ref{fig:2}: 
the symbols identify the different 
asymptotic states, while the curves correspond to the marginal stability lines 
(determined theoretically, see Fig.~\ref{fig4}). 
The simulation of Eqs.~(\ref{eq2},\ref{eq1}) was performed for $N=1000$; 
larger ensemble-size have been considered for several points without 
observing any essential 
difference. The transient time was $2\cdot 10^4$ time units.

\begin{figure}
\begin{center}
\includegraphics[width=0.48\textwidth,clip=true]{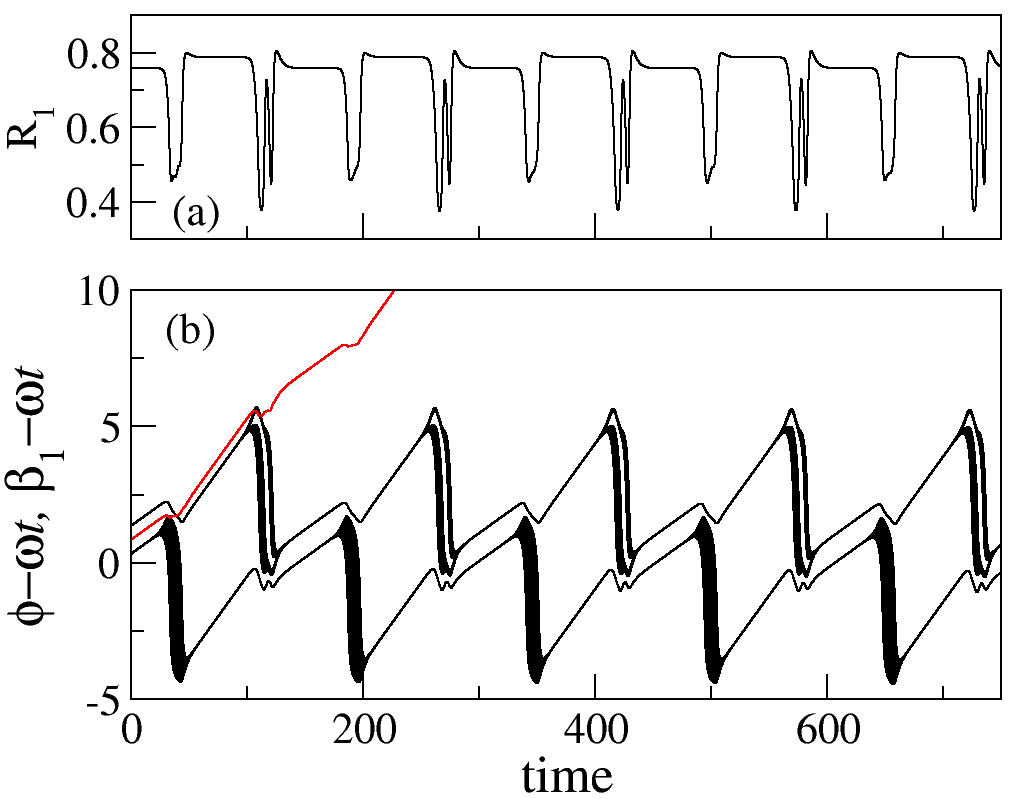}
\caption{Dynamics of the heteroclinic cycle. (a) Time variation 
of $R_1$ reflects the so-called slow switching.
(b) Evolution of the phases of all oscillators (black) and of the mean field (red) 
- after subtracting the average growth.
For rather long epochs oscillators are grouped into two clusters with almost identical 
phases. However, these states are unstable and alternately each group widens 
until its width becomes of order one, and then shrinks again. 
Notice, that there is no transfer of elements between the two groups,
but the mean field frequency is larger than that of oscillators.
Indeed, we see that in the coordinate frame co-rotating with the frequency 
$\w$, the phase $\beta_1$ drifts away.
Parameters are $\gamma_2=\pi$, $\gamma_1=1.35$, $N=1000$ (the dynamics is preserved
for the ensemble size as large as $N=5000$). 
The initial conditions are a perturbed splay state. Size of the two groups 
is close but not equal to $N/2$. 
A movie showing the time evolution of this regime is included in the supplementary data
(see supplementary movie 1). 
}
\label{fig:hetero}
\end{center}
\end{figure}

In order to avoid the spurious formation of clusters in the HC states,
we made the oscillators slightly heterogeneous. Namely, their frequencies 
(all equal to zero in  Eqs.~(\ref{eq2},\ref{eq1})) have been taken as uniformly 
distributed in the interval $[-0.5\cdot10^{-12},0.5\cdot10^{-12}]$. 
This diversity, which is crucial for the detection of HCs,
had no influence on the other dynamical states.
It has been checked that variation of the inhomogeneity in the range 
$10^{-10}-10^{-14}$ has no essential effect on the parameters of the 
heteroclinic cycle.
For an automatic detection of the states, all oscillators characterized by phase
differences smaller than $10^{-8}$ have been identified as belonging to the
same cluster (this threshold was chosen by trial and error, after a visual 
inspection of the observed regimes). By monitoring the number of clusters,
we have found that their number varies in time only in the case of HCs. 
This is due to the fact that the two cluster-widths greatly change over time,
as illustrated in Fig.~\ref{fig:hetero}b, where the phase of each oscillator
is plotted versus time in a co-rotating frame. There we see that time intervals
where the cluster amplitude is of order one alternate with relatively long
periods where the width is extremely small, thus yielding spuriously detected
clusters. The asymmetry between the behavior of the two clusters
(one of them splits into two parts) follows from the non perfectly equal
size of the two clusters.
The periodic variation of the cluster-width is reflected
in periodic variation of the order parameter $R_1$ (Fig.~\ref{fig:hetero}a).
Furthermore, the average frequency of the mean field 
is larger than that of oscillators, as can be appreciated from the plot 
of the mean-field phase (Fig.~\ref{fig:hetero}b);
this difference emerges due to a permanent interchange 
of the leading and lagging clusters, discussed in the previous Section. 
Thus, the HC state
can be interpreted as a special form of quasiperiodic SCPS dynamics, 
with a smooth but non-stationary distribution.  
A movie showing the time evolution of heteroclinic cycles is included in the supplementary data
(see supplementary movie 1). 

Furthermore, we noticed that some regimes are approached after a very long transient.
This is particularly true close to the stability border of different states.
The points that appear as SCPS for $\gamma_2\approx 4.75$, 
$\gamma_1\approx \pi/2$, converge to two-cluster states if the transient time
is increased.
Next, consider the thin ``belt'' of SCPS states close to the theoretical curve, 
which denotes the border of stability of full synchrony.
Computations show that upon increasing the transient time, some points that appear
as SCPS states in Fig.~\ref{fig:2} progressively converge towards an HC.
Thus, the SCPS in the ``belt'' domain is probably a very long transient regime. 
However, for the points in the main domain of SCPS, e.g. 
for $\gamma_2=\pi$, $\gamma_1=1.5$, 
the partially synchronous dynamics seems to be the asymptotic state, 
at least it survives for $10^7$ time units. 

Further states have been occasionally detected 
(see the cyan symbol in Fig.~\ref{fig:2}).
In particular three-cluster states are found for $\gamma_2=4.75$ 
and $1.58\le \gamma_1\le 1.66$, 
where the degree of synchronization may even oscillate in time (three-cluster states
had been already observed in \cite{Hansel-Mato-Meunier-93} 
for a slightly different amplitude of the second 
harmonic).

\begin{figure}
\begin{center}
\includegraphics[width=0.6\textwidth,clip=true]{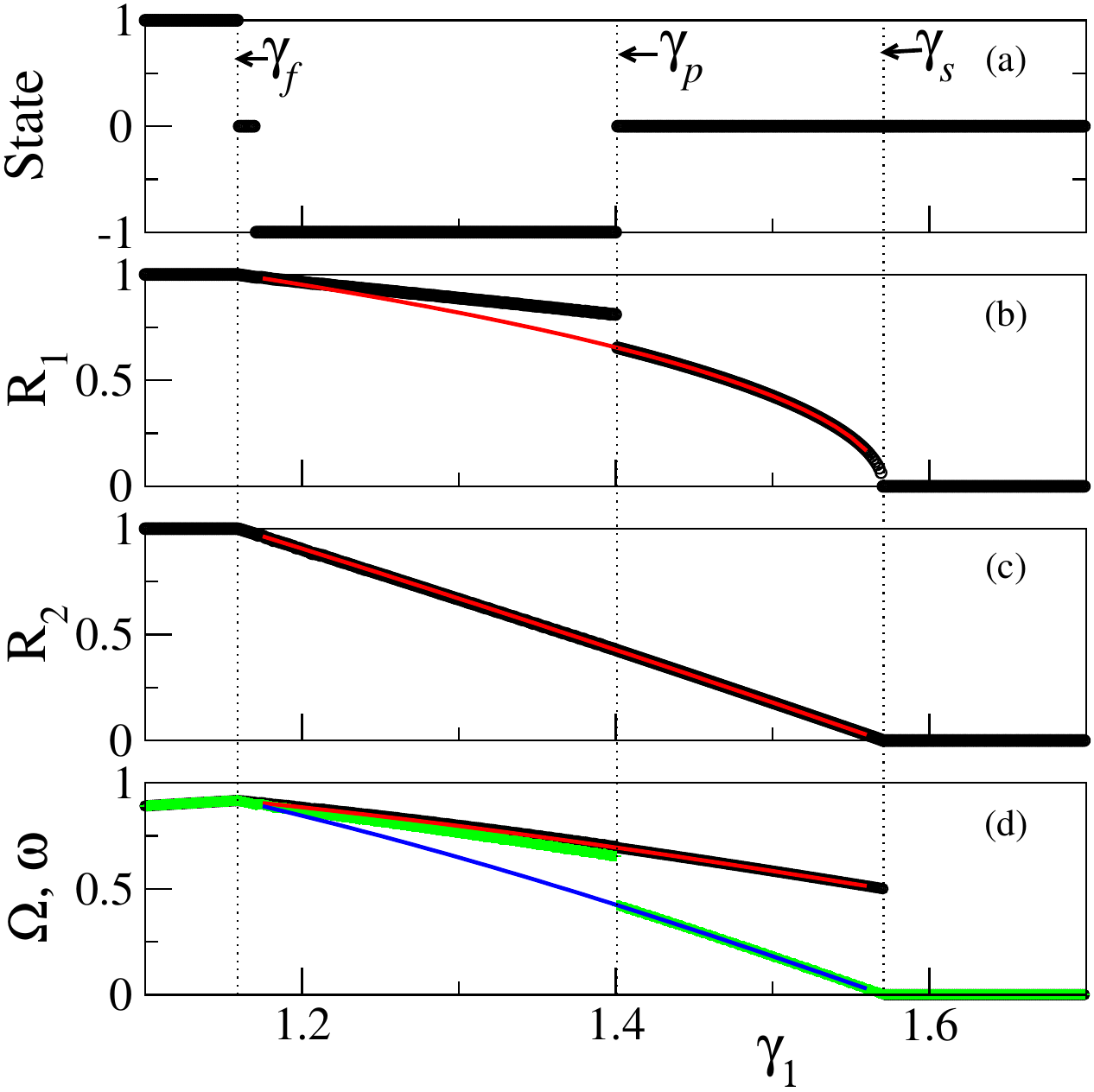}
\caption{Numerical results for $\gamma_2=\pi$ upon varying $\gamma_1$. 
The initial conditions for each point are perturbed splay states
(one oscillator-phase being slightly displaced).
Panel (a) shows the resulting state after a transient $5\cdot10^4$, coded by an integer.
Heteroclinic cycles are coded by $-1$; for the other states, the code equals the 
number of clusters.
Panels (b) and (c) exhibit the first and the second order parameters, 
respectively. Black circles correspond to numerical simulations;
red lines are the results obtained by solving Eq.~(\ref{eq:partial_ana}).
Macroscopic and microscopic frequencies are shown in (d): 
Black circles and green squares are the frequencies of the mean field
and of the oscillators, respectively, determined numerically 
(since the oscillators are slightly 
non-identical, the oscillator frequency is obtained by averaging over all units).
Red and blue curves correspond to the analytical solution.
The vertical dotted lines mark $\gamma_f$, $\gamma_p$, and $\gamma_s$, respectively
(see text). 
}
\label{fig:3}
\end{center}
\end{figure}

A detailed quantitative analysis has been performed along the line $\gamma_2 = \pi$ in
the parameter space. The results are shown in Fig.~\ref{fig:3}, where 
one can recognize three critical points:
(i) $\gamma_s=\pi/2$ signals the loss of stability of the splay state;
(ii) $\gamma_f = \arccos(2a)\approx 1.159$ signals the loss of stability of the fully
synchronous state;
(iii) $\gamma_p\approx 1.401$ signals the loss of stability of SCPS.

Upon decreasing $\gamma_1$ from $\gamma_s$, SCPS is first born through a
Hopf bifurcation from the splay state;
$R_1$ and $R_2$ become strictly larger than zero and a finite
mean-field frequency $\W$, which corresponds to the frequency of the 
Hopf bifurcation, appears discontinuously. 
Simultaneously, the microscopic
frequency $\omega$ deviates from zero because of the macroscopic modulation,
without revealing any locking with $\W$. 

Interestingly, $\W$ is positive and always larger than $\omega$: 
this is at variance with the scenario observed 
in~\cite{Politi-Rosenblum-15}, where the opposite was found. 
However, we should also recall that an equivalent scenario is observed in
the upper eyelet, once the transformation $\phi\to-\phi$ has been performed. 
In fact, such a change of variable would lead to the scenario observed in LIF neurons. 
Notice that for Kuramoto-like model of nonlinearly coupled oscillators 
\cite{Rosenblum-Pikovsky-07,Pikovsky-Rosenblum-09} both cases ($\W>\w$ and 
$\W<\w$) are possible.
Altogether, the existence of a finite difference between microsocpic and 
macroscopic frequencies is a key signature of a quasiperiodic SCPS.

Upon further decreasing $\gamma_1$, we enter a region where the dynamics 
converges to HCs which are characterized by a sudden increase of the (average) 
$R_1$, while no discontinuity is exhibited by
$R_2(\gamma_1)$. When, finally, $\gamma_f$ is approached, unsurprisingly 
$R_1$ and $R_2$ converge to 1, while both $\W$ and $\omega$ converge to the 
frequency of the fully synchronous state.
The nature of the bifurcation is not clear.
We briefly comment in the next section, while discussing the stability of SCPS.

\begin{figure}
\begin{center}
\includegraphics[width=0.55\textwidth,clip=true]{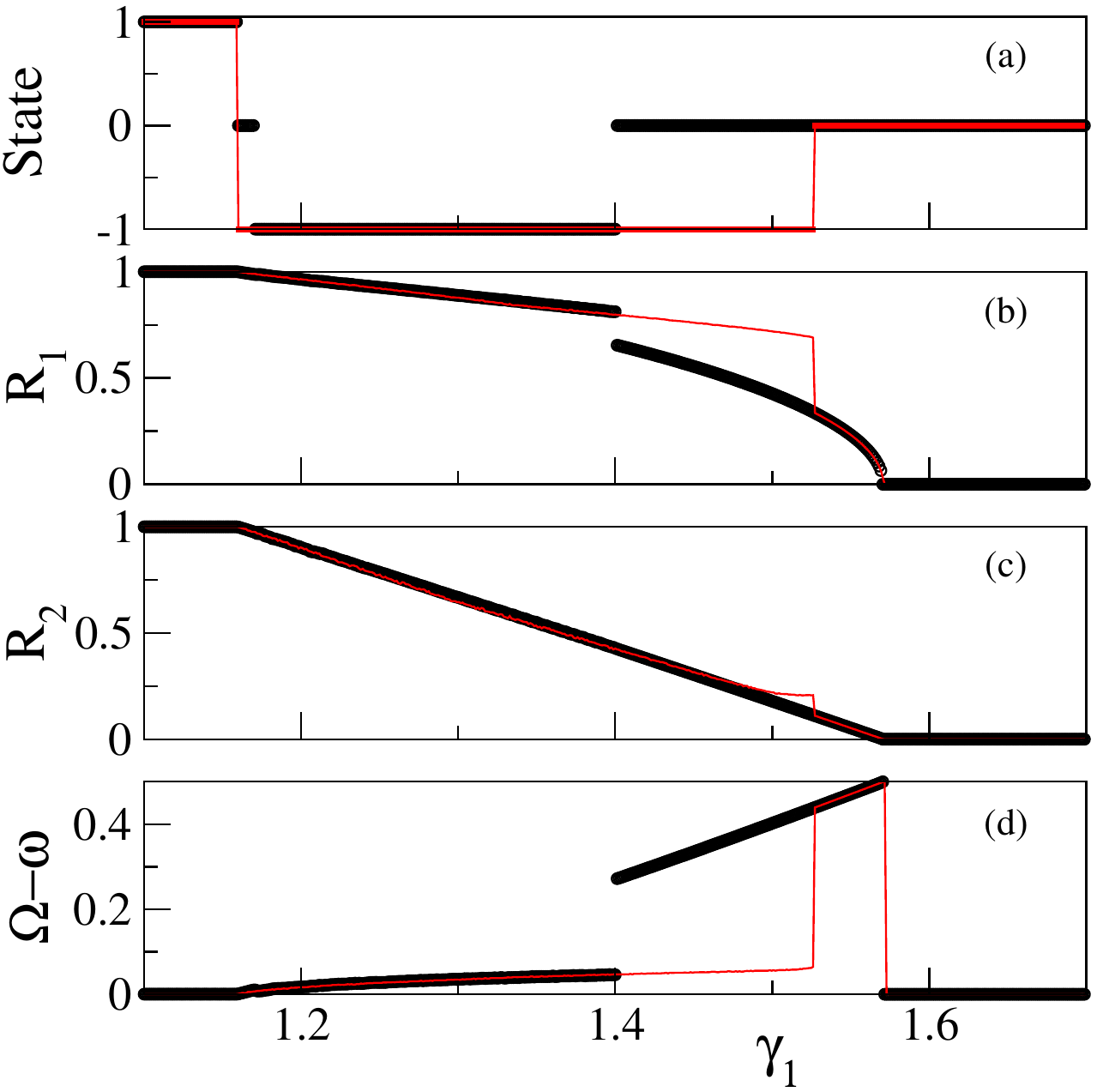}
\caption{Illustration of the multistability in the collective dynamics,
for $\gamma_2=\pi$ upon varying $\gamma_1$. 
Black symbols correspond to perturbed splay initial conditions 
(see also Fig.~\ref{fig:3}); red curves correspond to 
slow increase of $\gamma_1$.
Panels (a-c) are similar to those in Fig.~\ref{fig:3}, panel (d) 
shows the frequency difference. Computations with slow decrease of 
$\gamma_1$ are not shown because their results coincide with 
black symbols. 
}
\label{fig:multi}
\end{center}
\end{figure}

The sudden jump observed for $\gamma_1=\gamma_p$ suggests the possible existence of
multistability. Accordingly, we have performed two additional series of simulations,
by progressively increasing (decreasing) $\gamma_1$, and choosing the
new initial condition as a slightly perturbed version of the final
configuration for the previous value of $\gamma_1$.
As shown in Fig.~\ref{fig:multi}, a bistability region is indeed found
where HCs and SCPS are simultaneously stable.

\subsection{Macroscopic analysis}

For a more detailed characterization of SCPS, we have determined the corresponding 
probability distribution by solving Eq.~(\ref{eq:partial_ana}), as discussed in 
section~\ref{sec:PS}. This method allows to obtain 
the phase distribution even when it is unstable.
As it can be seen in Fig.~\ref{fig:3} (see the thin solid lines), the results are 
consistent with the direct numerical simulations wherever SCPS is stable. 
Moreover, it is found that SCPS exists
also in the interval $[\gamma_f,\gamma_p]$ and it reconnects to the fully 
synchronous state.
As for the microscopic frequency $\omega$, given by
\[
\omega = \W+2\pi \eta \; .
\]
it also agrees with the numerical simulations.

\begin{figure}
\begin{center}
\includegraphics[width=0.45\textwidth,clip=true]{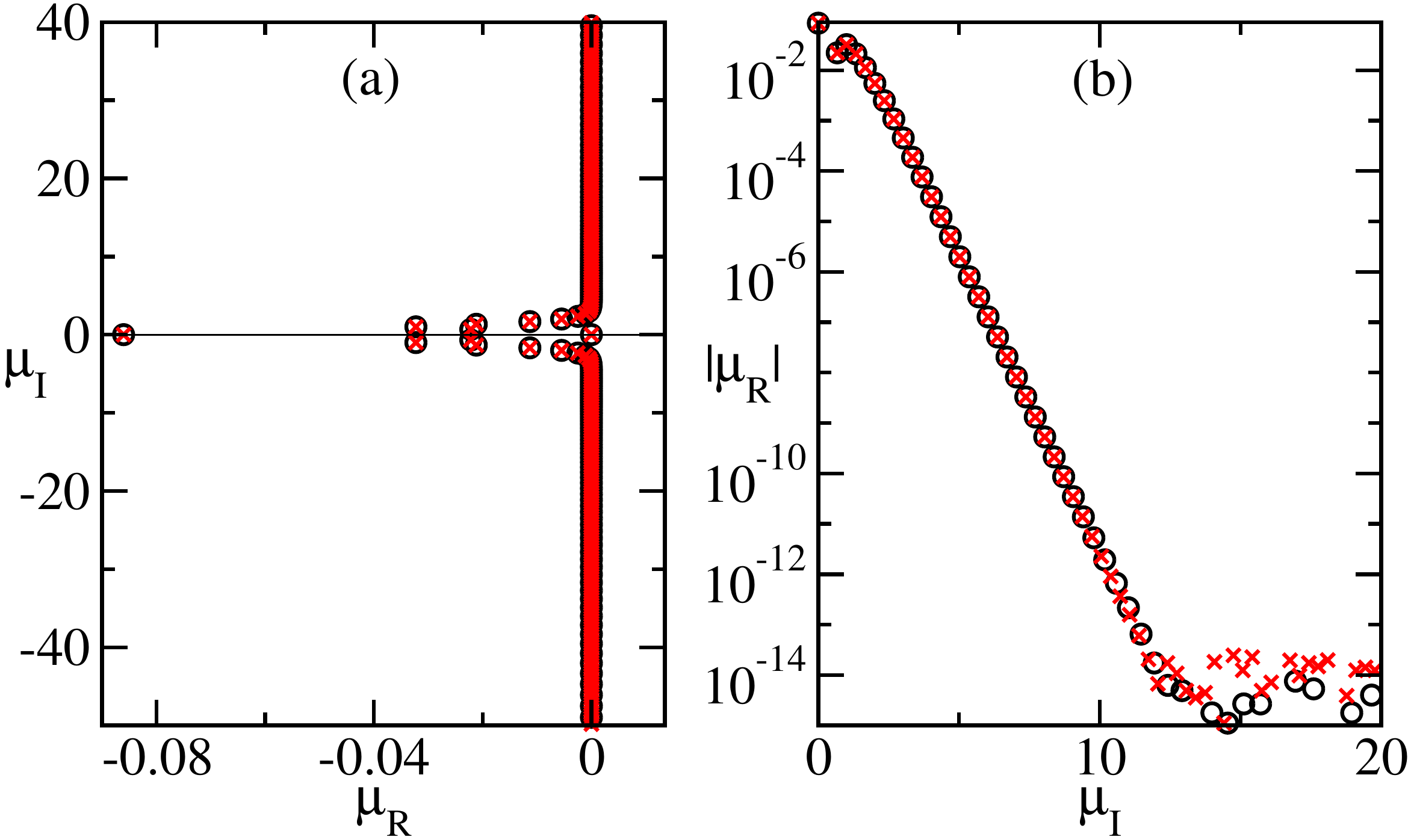}
\caption{
Eigenvalues of SCPS for $\gamma_2=\pi$ and $\gamma_1 = 1.45$. 
Circles and crosses correspond
to a truncation after 100 and 500 modes, respectively.
} 
\label{fig:eigen}
\end{center}
\end{figure}

The stability analysis carried out in Sec. III allows determining $\gamma_p$ by solving the
eigenvalue problem (\ref{eq:lineargen}). A typical spectrum is plotted in Fig.~\ref{fig:eigen}.
There we see that, with a few exceptions, the eigenvalues tend to align along the imaginary axis. 
Besides the zero associated to the conservation of the total probability,
there exists a second zero eigenvalue
which follows from the invariance under phase-translation of the probability density.
In panel (b) we see that the real part of the eigenvalues decreases exponentially with their imaginary component.
The plateau seen for large $\mu_I$ is a consequence of the finite numerical accuracy of the simulations. 
Interestingly, finite-size effects are practically absent (at least in this parameter region): a
larger number of Fourier modes leads to  eigenvalues with a larger frequency (imaginary component) and
an exponentially small real part.

\begin{figure}
\begin{center}
\includegraphics[width=0.4\textwidth,clip=true]{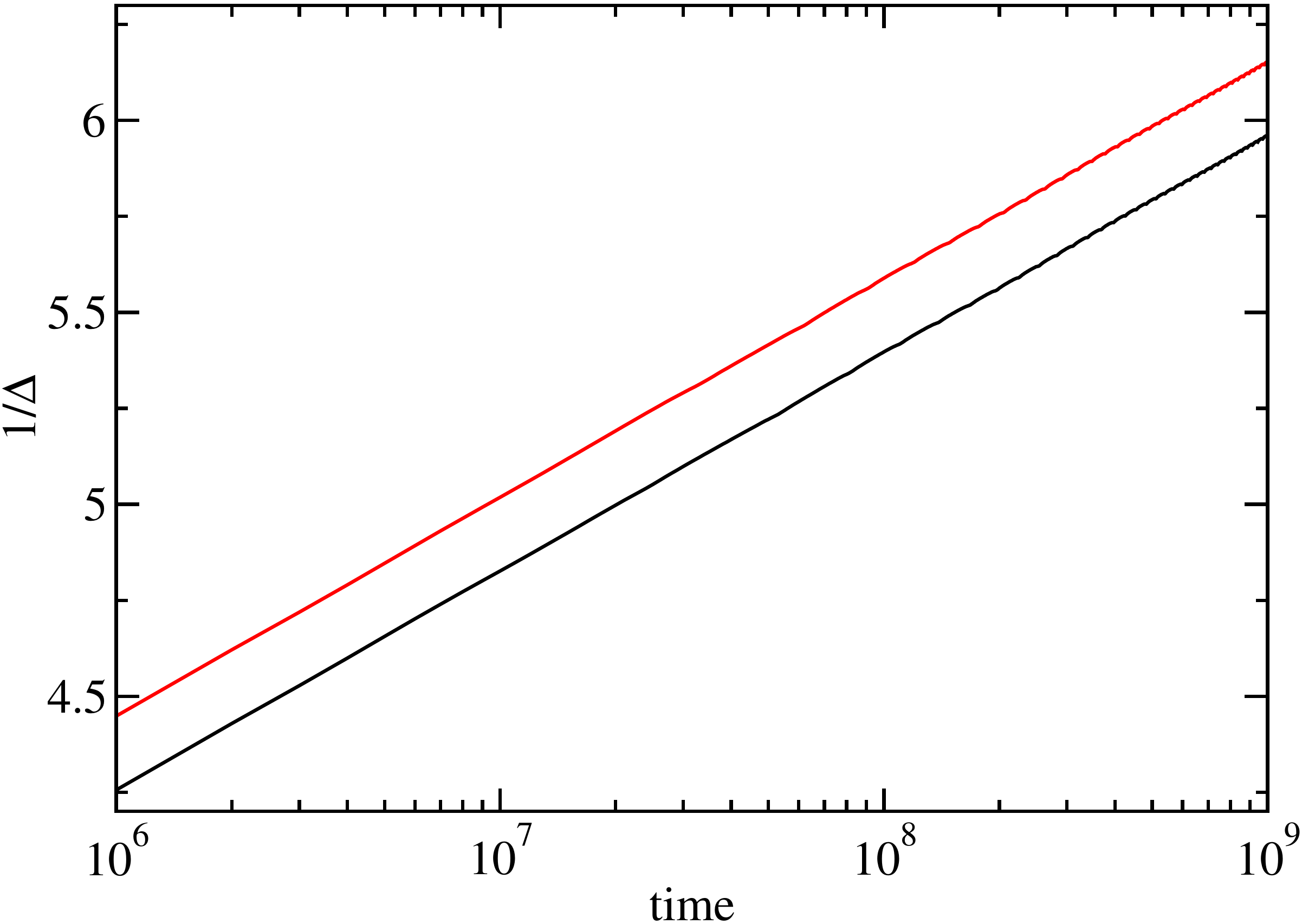}
\caption{
Evolution of an initially flat distribution with a gap $\Delta(0)$ for
$\gamma_1=1.5$ and $\gamma_2=\pi$, where SCPS is stable.
The inverse gap-width $1/\Delta$ is shown as a function of time,
for $\Delta(0) =\pi/7$ and $\pi/5$, (black and red lines, respectively).}
\label{fig:conver}
\end{center}
\end{figure}

In practice, SCPS is marginally stable, as there is an infinite number of eigenvalues 
with a practically vanishing real part. The evolution of a uniform distribution
initially confined to an interval of size $2\pi-\Delta_0$ offers the chance to 
appreciate the role of the weakly attracting directions. As seen in
Fig.~\ref{fig:conver}, the gap size goes to zero, but it does so in an
extremely slow way, namely as  $a/\ln t$.

In the past, the evidence of a similarly slow convergence was found for the splay state
itself. In the context of pulse-coupled oscillators with an analytic velocity field
a similar set of exponentially decreasing real parts had been 
observed~\cite{Calamai-09,Olmi-14}. 
In the context of Kuramoto-Daido models, the strength of the real part is
directly proportional to the amplitude of the Fourier component of the
coupling function (see Eq.~(C2) in \cite{Politi-Rosenblum-15}), so that, when
the number of Fourier modes is finite, infinitely many
strictly marginal directions are present (in the thermodynamic limit). 
This is reminiscent of the Watanabe-Strogatz theorem 
\cite{Watanabe-Strogatz-93,Watanabe-Strogatz-94}
which implies that 
(for a strictly mono-harmonic coupling function) infinitely many directions
are not only linearly marginally stable but actually correspond
to conservation laws.
Our results show that the existence of conservation laws breaks down
already when two harmonics are considered. In fact, although exactly
marginally stable directions are detected in the analysis of the splay state
(here only two Fourier modes are present, so that only two directions can
be strictly (un)stable), the same is no longer
true for SCPS, as all modes have a weak but finite stability.

\begin{figure}
\begin{center}
\includegraphics[width=0.75\textwidth,clip=true]{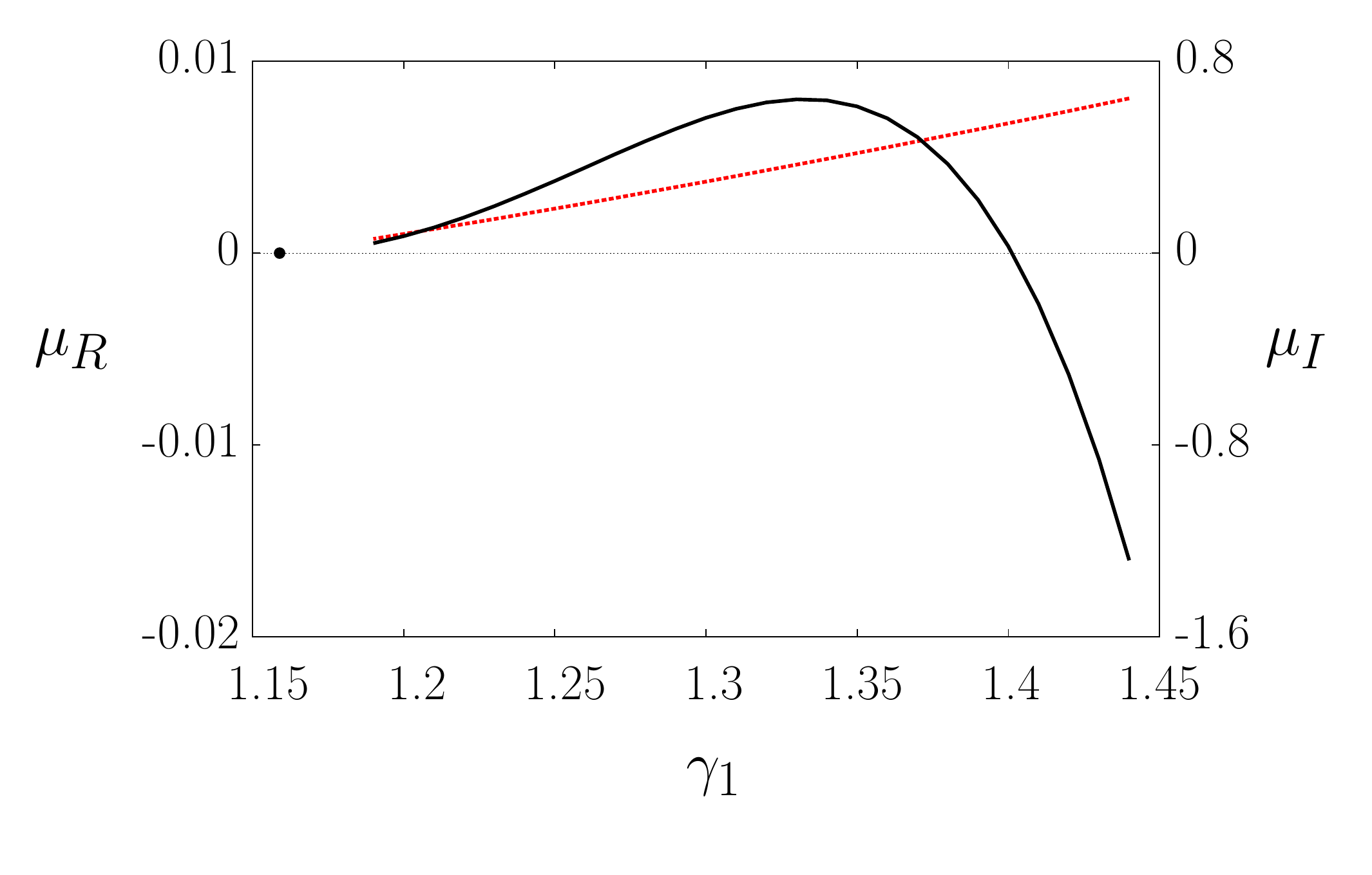}
\caption{Stability analysis of the splay state for the biharmonic model.
Real and imaginary part of the eigenvalues for the maximally unstable direction for $\gamma_2=\pi$ 
are shown versus $\gamma_1$. The left (right) scale refers to the real (imaginary) part. 
The solid and dashed lines correspond to the real and imaginary
components, respectively. The full circle identifies the point where the 
fully synchronous solution changes stability.}
\label{fig:maxeig}
\end{center}
\end{figure}

In Fig.~\ref{fig:maxeig} we plot the real and imaginary part of the most 
unstable eigenvalue 
versus $\gamma_1$.  The data confirms that linear instability occurs below $\gamma_p$: 
the bifurcation is of Hopf type.
SCPS is maximally unstable around $\gamma_1\approx 1.33$. 
By further decreasing $\gamma_1$, both
the real and the imaginary parts decrease to zero, while approaching $\gamma_f$. 
From the dependence of
the real part, it seems that the scaling behavior is quadratic in the distance from the critical point.
The nature of the bifurcation is unclear: simulations close to $\gamma_f$ are not reliable.
It is nevertheless instructive to notice that the weak instability is consistent with the observation
of partial synchrony over extremely long time scales mentioned in the beginning of this section.
The correct identification of the critical point is further confirmed in Fig.~\ref{fig:2}, where the
triangles are the outcome of the stability analysis for three different choices of $\gamma_2$.

\section{Other setups}

So far, we have discussed the occurrence of SCPS in a simple Kuramoto-Daido setup where
the coupling function is composed of just two Fourier harmonics. 
The Kuramoto-Daido representation generally holds in the weak coupling limit; 
however, the bi-harmonic coupling function may be
too simple a model and it is therefore worth comparing with other setups. 
As recalled in the introduction, SCPS was first observed in pulse-coupled 
LIF neurons~\cite{vanVreeswijk-96}. 
In Ref.~\cite{Politi-Rosenblum-15}, it has been shown that the system can be 
effectively described by a Kuramoto-Daido model with a coupling function containing
several nonnegligible harmonics. 
In the first subsection we show that such harmonics contribute to destabilizing
the two-cluster states that are, in fact, never observed.
In the second subsection we discuss an ensemble of two-dimensional Rayleigh
oscillators esch described by two variables, showing that SCPS can be generated
in this case as well. A phase reduction works also in this latter case, where
the higher harmonics play a different role:
they are responsible for the destabilization of the splay state, giving rise to more
structured cluster states.

\subsection{LIF neurons}

One of the important open questions in the study of ensembles of phase-oscillators 
is that of determining \textit{a priori} whether a given coupling function $G$ 
gives rise to either smooth distributions, or macroscopic clusters,
or both. In this perspective it is instructive to explore the difference between 
the scenario seen in the 
biharmonic model and that observed in an ensemble of LIF neurons. 
In such a case the model can be reduced
to a Kuramoto-Daido setup with
\begin{equation}
\label{eq:Gphi}
G(\vp) = g_1(g_2-1+\vp)\mathrm{e}^{\alpha(1-\vp)/\nu}+g_3\mathrm{e}^{(1-\vp)/\nu}-g_4 \; .
\end{equation}
(see Ref.~\cite{Politi-Rosenblum-15} for a precise definition of the various 
parameters - notice that here we are using a different notation -- $\vp$ instead of $\phi$ --
since in the above equation the phase is normalized between 0 and 1).
Correspondingly we change notations in this paragraph.
Straightforward calculations reveal that two-cluster states exist
also in the above model, the main difference being, however, that 
now (for $\alpha=6$), the average exponent $\lambda_a > 0$,  
so that two-cluster states are effectively unstable, 
i.e. no any spurious cluster is appearing due to the finite precision of computations.
This explains why they have never been seen in numerical simulations.

It is natural to ask whether this holds true for arbitrarily large $\alpha$, when the
SCPS becomes increasingly close to full synchrony.
We proceed by expanding $G_A$ around 0 in the limit of large $\alpha$.
One first finds
\[
G'_A(0) = - \frac{\nu}{a+g\nu}\left( \mathrm{e}^\tau-1 \right)\;,
\]
which is negative and finite. Comparison with Eq.~(\ref{eq:cluster2}),
implies that the synchronous solution is always unstable.
As for the second derivative, the leading order
\[
G''_A(0) = \alpha^3 \frac{\mathrm{e}^\tau -1}{2\nu(a+g\nu)}
\]
is positive and increasingly large.
As a result, on the basis of the first two polynomial terms of $G_A(\delta)$, 
one finds that it vanishes also for
\begin{equation}
\delta_0 = \frac{2\nu^2}{\alpha^3} \;.
\end{equation}
This phase shift identifies a two-cluster state;
its value decreases as the cubic power of the 
pulse-width (equal to $1/\alpha$).
Under this quadratic approximation, the slope of $G_A$ in 0 and $\delta_0$ are equal and opposite
to one another, so that $\lambda_T =0$. This marginal value requires going one order beyond in the 
perturbation analysis. The computation of the third derivative shows that it is negative
and of order $\alpha^4$.
As a result its contribution to the
derivative in $\delta_0$ is of order $\alpha^4 \delta_0^2 \approx 1/\alpha^2$. This makes
the sum of the derivatives in 0 and $\delta_0$ slightly negative and proves that the
two-cluster state is always effectively unstable.

If we recall the definition of $G_A$, one might be surprised to see that its expansion
around 0 contains the second, even, power of $\varphi$. One should however remember that the
original function $G$ is continuous but not even of class $\mathcal{C}^{(1)}$;
so it is not unnatural to expect a discontinuity of some derivative in zero.

In smooth models, in the vicinity of the bifurcation where the synchronized solution loses its
stability (e.g., for the biharmonic model) one can write
\begin{equation}
G_A(\varphi) = -\varepsilon \varphi + A \varphi^3\;.
\end{equation}
If $A>0$ and $\varepsilon>0$, then there exists a second zero
$\delta_0 = \sqrt{\varepsilon/A}$, which corresponds to the clustered solution.
Therefore
\begin{equation}
\lambda_T = 2 \varepsilon - 3 A \varphi_0^2 = -\varepsilon <0 \; ,
\end{equation}
so that smooth interaction functions necessarily lead to effectively stable two-cluster
states (at least with respect to intracluster perturbations).

Even more, Eq.~(\ref{eq:intra_biha}) yields that in the biharmonic model
no effectively unstable cluster may exist: only heteroclinic cycles.
For these clusters to exist, higher harmonics are needed.

\subsection{Rayleigh oscillators}

In order to provide further evidence on the ubiquity of SCPS,
we present a numerical study of globally coupled identical Rayleigh oscillators
\footnote{This system is equivalent to the van der Pol equation;
they are related via the variable substitution 
$\dot x \to x/\sqrt{3}$. We prefer this formulation of the model for technical reasons.}.
The equations are 
\begin{equation}
 \ddot x_k-\zeta(1-\dot x_k^2)\dot x_k + \omega^2x_k=
 \e {\rm Re}\left[e^{i\gamma}(X+i Y)\right ] \;,
 \label{rayens}
\end{equation}
where $X=N^{-1}\sum_kx_k$ and $Y=N^{-1}\sum_k\dot x_k$ are two mean fields, while
$\e$ is the coupling strength. 
Finally, the control parameter $\gamma$ accounts for a phase shift 
of the coupling term: it determines whether the interaction is attractive or repulsive.

It is well-known that uncoupled units in Eqs.~(\ref{rayens}) exhibit 
limit-cycle oscillations, 
while the nonlinearity parameter $\zeta$ determines 
the stability of the limit cycle. 
Below we consider $\zeta=5$; for this parameter the transversal Lyapunov exponent is -7.358.
Therefore, adiabatic elimination of the amplitude is rather meaningful. 

An appropriate order parameter is
\begin{equation}
\rho=\mbox{rms}(X)/\mbox{rms}(x)\;,
\label{rayopar}
\end{equation}
where ``rms'' means root-mean-square of the time evolution. 
The splay state is characterized by a constant mean field and thereby $\rho=0$. 
In the fully synchronous state, for identical oscillators, 
the microscopic and macroscopic
dynamics are equivalent to one another so that $\rho =1$.

\begin{figure}
\begin{center}
\includegraphics[width=0.6\textwidth,clip=true]{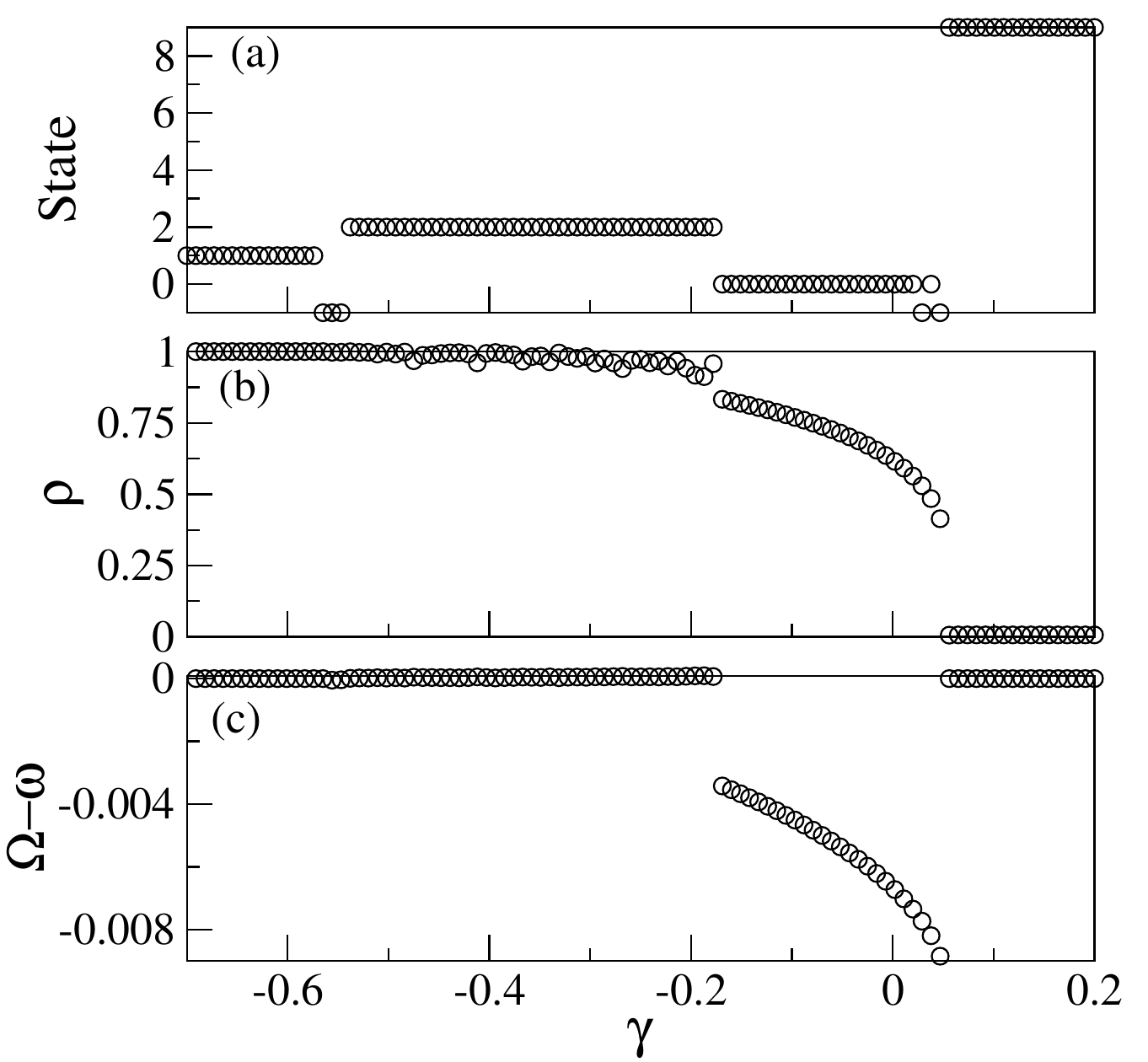}
\caption{Self consistent partial synchrony is observed in model (\ref{rayens})
for a rather broad range of $\gamma$ values. 
Panel (a) shows the final state: the $y$ variable corresponds to the number of clusters, 
if it does not vary within a long time interval, or to $-1$, otherwise 
(such a regime appears at the border between the synchronous and two-cluster
states and between SCPS and nine-cluster states; possibly these states simply
suggest the presence of very long transients).
Thus, the state coded by zero corresponds to SCPS. 
Panel (b) shows the order parameter. In the two-cluster states it is almost one, 
because the clusters are close to each other. In the SCPS state it varies between
$\approx 0.4$ and $\approx 0.8$.
Finally, panel (c) shows the frequency difference between microscopic and macroscopic
dynamics, which differs from zero in SCPS.  
}
\label{fig:ray2}
\end{center}
\end{figure}

The scenario resulting for $N=1000$, $\e=0.05$,  
$\zeta=5$, and transient time $7.5\cdot10^{5}$ is reported in
Fig.~\ref{fig:ray2}. 
It is reminiscent of that observed in the biharmonic
model, with some differences.
SCPS establishes itself in between the parameter range where full synchrony
is stable (below $\gamma \approx-0.6$) and the region where $\rho$ is
negligible (above $\gamma \approx 0.05$). In this case, the mean field is 
slower than the individual oscillators, as in the upper eyelet of the
biharmonic model, cf.~Fig.~\ref{fig:2}.
The regime characterized by a vanishing $\rho$ is not asynchronous but a 
symmetric 9-cluster state (see also below).
In fact, the splay state turns out to be unstable in the full range of 
$\gamma$ values that we have explored.
Finally, in the interval $[-0.6,-0.2]$ we see a slow convergence towards a
two-cluster state  (the large value of the order parameter $\rho$ is due to the
closeness between the two clusters).

The dynamics of the coupled system Eq.~(\ref{rayens}) can be better understood by 
performing a phase reduction. 
This can be done by introducing a phase variable
for each individual oscillator, making reference to the uncoupled limit
(i.e. $\e=0$): $\phi_k=2\pi t_k/T$, where $t_k$ is the time elapsed from
the passage through a chosen origin $x=0$, $\dot x>0$, 
while $T$ is the oscillation period. 

\begin{figure}
\begin{center}
\includegraphics[width=0.7\textwidth,clip=true]{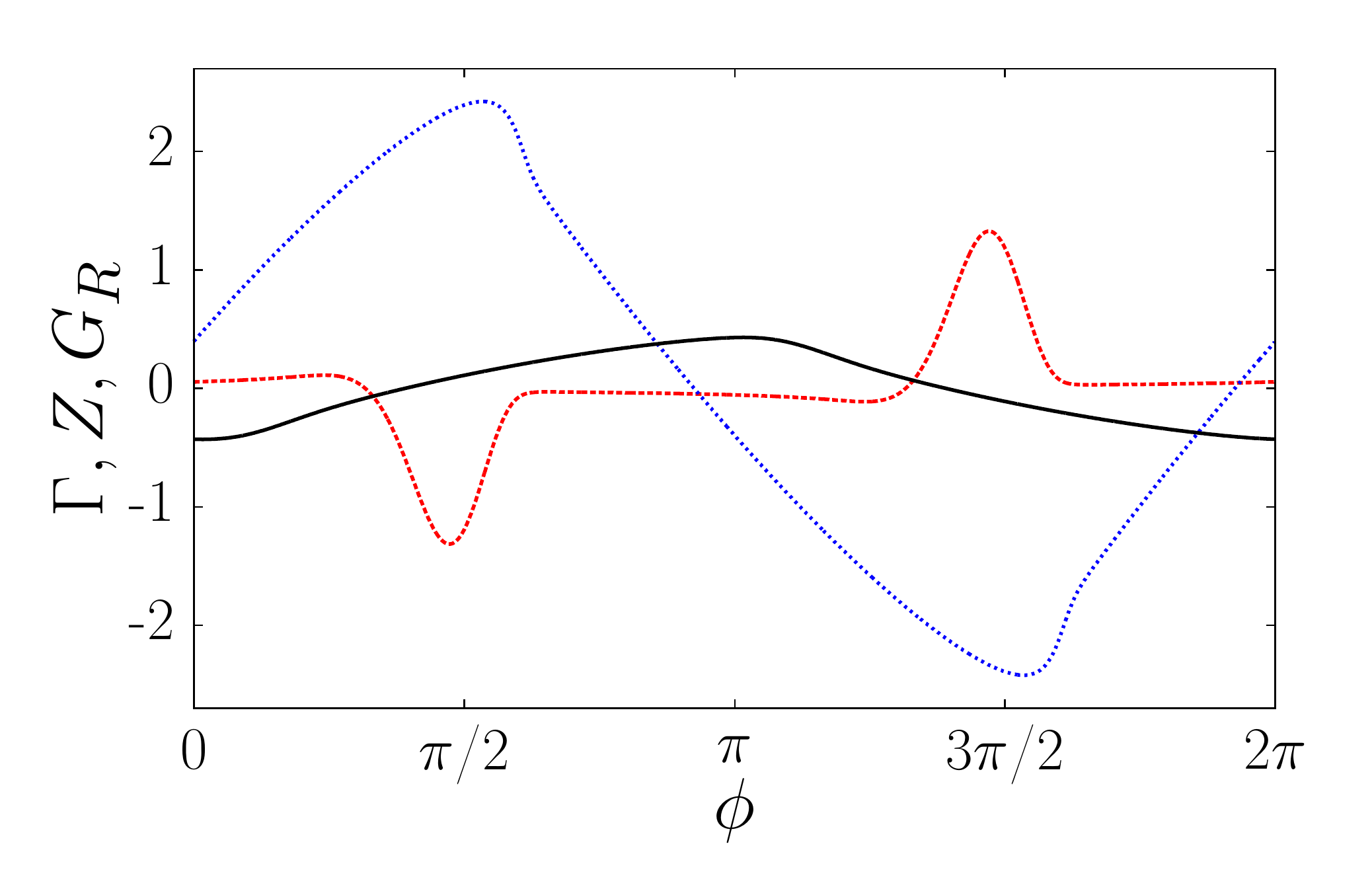}
\caption{Phase response curve $\Gamma$ (red dashed line), the forcing 
function $Z$ (blue dotted) of the Winfree model Eq.~(\ref{rayens2}), 
and the coupling function $G_R$ of the Kuramoto-Daido reduction (black) for $\gamma=-0.4$.
}
\label{fig:ray3}
\end{center}
\end{figure}

In the weak coupling limit, the original Rayleigh oscillators (\ref{rayens}) can be mapped
onto a Winfree model
\begin{equation}
\dot \phi_k = \omega_0 + \frac{\varepsilon}{N} \Gamma(\phi_k) \sum_j Z(\phi_j) \; ,
 \label{rayens2}
\end{equation}
where $\Gamma(\phi)$ is the phase response curve (PRC), while $Z(\phi)$ is the forcing function.
$\Gamma(\phi)$ can be obtained by following a standard approach: 
it corresponds to the phase shift imposed by an infinitesimal 
kick when the phase of the oscillator is $\phi$. The resulting PRC is reported in
Fig.~\ref{fig:ray3} (see the red dotted curve). 
The forcing function $Z$ is instead
obtained by expressing the coupling term due to the $j$th oscillator 
${\rm Re}\left[\mathrm{e}^{i\gamma}(x_j+i \dot x_j)\right ]$ as a function of $\phi_j$
(see the blue dotted curve in  Fig.~\ref{fig:ray3}).
Finally, following Refs.~\cite{Golomb-Hansel-Mato-01,Politi-Rosenblum-15},
one can further map the Winfree model (\ref{rayens2})
onto a Kuramoto-Daido model, by computing the convolution of $\Gamma$ and $Z$, i.e.
\begin{equation}
G_R(\phi) = \int \Gamma(\phi-\psi) Z(\psi)d\psi \;.
\end{equation}
The resulting coupling function corresponds to the black curve in Fig.~\ref{fig:ray3}. 
Fourier analysis shows that even modes are absent: this follows from the symmetry of
the limit cycle: adding $\pi$ to the phase results in changing the sign of both
the PRC and forcing term.

In order to test the validity of the Kuramoto-Daido reduction, it has been simulated for
eight harmonics.
The resulting scenario is in close agreement with that one exhibited by the original ensemble of
Rayleigh oscillators, including the observation of 9-cluster states.
Within the Kuramoto-Daido representation, one can easily perform a stability
analysis of both the splay and synchronous state.
As discussed in Ref.~\cite{Politi-Rosenblum-15}, the stability of the splay
state is determined by the imaginary components of the Fourier modes of $G_R$. 
In Tab.~\ref{tab:ray1} we show the contribution of the first eight nonvanishing components
for $\gamma =0.2$, where a zero order parameter is observed. In fact, the splay state
turns out to be unstable because of the contribution of the 7th and higher harmonics.
This is at variance with the biharmonic model, where the stability was determined
by the first Fourier mode.

As for the fully synchronous state, its stability is determined by the sign of $G_R'(0)$.
The bifurcation point where it changes stability is $\gamma\simeq -0.573$
in agreement with the numerical simulations of the original model (\ref{rayens}).

For larger values of $\gamma$ and below -0.2, the Kuramoto-Daido model possesses
two-cluster states.  Following the analysis outlined in Sec.~\ref{sec:clusters},
we conclude that two-cluster states are effectively stable.
In fact, for $\gamma=-0.4$, the inter-cluster exponent $\lambda_I= -G_A'(\delta_0)=-0.0669 $
is negative, while $\lambda_E^+ = 0.0869$, and $\lambda_{E}^-=-0.1146$.
Accordingly the average exponent $\lambda_a = -0.01385$ is negative,
in spite of one of the two intra-cluster exponents being positive.
Altogether, the scenario is similar to that observed in the biharmonic model.
\begin{table}
\begin{center}
 \begin{tabular}{|c|c|c|}
 \hline
 Index & Real part & Imaginary part \\ \hline
1 &-1.6$ \times 10^{-2}$&  2.1$ \times 10^{-1}$\\
3 &  -5.6$ \times 10^{-2}$&  1.5$ \times 10^{-2}$\\
5 &  -1.6$ \times 10^{-2}$&  1.9$ \times 10^{-2}$\\
7 &   3.0$ \times 10^{-3}$&  1.0$ \times 10^{-2}$\\
9 &  4.0$ \times 10^{-3}$&  1.7$ \times 10^{-3}$\\
11 &  1.5$ \times 10^{-3}$&  5.0$ \times 10^{-4}$\\
13 &  3.8$ \times 10^{-4}$&  4.3$ \times 10^{-4}$\\
15 & 7.9$ \times 10^{-5}$&  1.8$ \times 10^{-4}$\\
\hline
\end{tabular}
\end{center}
\caption{Eigenvalues associated to the stability of the splay state for $\gamma=0.2$.
The index refers to the Fourier mode which they are associated with.}
\label{tab:ray1}
\end{table}

\section{Conclusions}

In this paper we have shown that self-consistent partial synchrony (SCPS) is
a general phenomenon, arising in many setups of globally coupled oscillators. 
This regime naturally emerges if the system is close to the border 
between full synchrony and asynchrony.
In fact, the minimal requirement for SCPS to arise is the presence of two harmonics in
the coupling function $G(\phi)$. This condition is naturally fulfilled in weakly coupled
oscillators away from the Hopf bifurcation. As an example we have indeed verified that
SCPS arises also in an ensemble of Rayleigh oscillators, where an approximate 
coupling function containing eight Fourier modes suffices 
to quantitatively reproduce the dynamics of the original model.
Altogether, SCPS is yet another dynamical regime that cannot be produced by 
the standard Kuramoto-Sakaguchi model. 
In fact, both partial synchrony with and without frequency difference can be obtained
in a model, 
based on strictly sinusoidal coupling function, but it requires a dependence 
on the coupling strength and/or the phase shift of the sine-function on the order
parameter, i.e. SCPS can be observed in case of nonlinear coupling only.  
This limitation disappears for our minimal Kuramoto-Daido model.

The mathematical structure of Kuramoto-Daido models allows for a semi-analytic 
treatment: it is not only possible to determine the probability distribution
of the phases, but also to perform a linear stability analysis and thereby 
determine the parameter range, where SCPS can be effectively observed.
In the biharmonic model and the Rayleigh oscillators, the loss of stability of SCPS 
drives the system towards a heteroclynic cycle, which can itself be interpreted as a 
(more structured) form of SCPS: 
here, besides a difference between the microscopic and mean field frequencies, 
a pulsation of the amplitude is present. In LIF neurons, instead, SCPS is always 
stable, while two-cluster states are always unstable.

It would be interesting to discover whether and under which conditions other kinds 
of bifurcation 
can drive SCPS towards more complex forms of collective dynamics. 
This question is related to that of identifying the number of relevant collective variables.
A fairly trivial answer can be given in the case of perfect clusters, 
as it boils down to studying low-dimensional
networks composed of a few ``supernodes'' (the clusters themselves).
The question is much less trivial in the context of smooth distributions
such as those associated to SCPS. 
Possibly, the first example of a complex behavior in a globally coupled 
partially synchronized system was given in \cite{Han-Kurrer-Kuramoto-95}, 
where the Morris-Lecar neuronal oscillators were analyzed numerically. 
However, this is a setup where phase-reduction is not globally possible.
More recently, evidence of a chaotic collective behavior has been found
in a population of quadratic integrate-and-fire neurons~\cite{Pazo-Montbrio-16},
where, however, the higher dynamical complexity is triggered by the presence of 
delayed interactions. So the question whether identical phase-oscillators can lead
to collective chaos is still open.

Another open general problem is that of using the information encoded in the coupling 
function to predict whether SCPS and/or cluster states can be generated. 
In the case of a sinusoidal coupling, the situation is 
simplified by the fact that clusters are not possible: their existence is excluded
by the Watanabe-Strogatz theory \cite{Watanabe-Strogatz-93,Watanabe-Strogatz-94}, 
see also \cite{Engelbrecht-Mirollo-14,Pikovsky-Rosenblum-15}. 
Thus, when the splay and synchronous states are both unstable, SCPS is the only 
possible solution.
In more general contexts cluster states sometimes coexist with SCPS, as well as with
chimera-like solutions.

\section*{Acknowledgement}
We wish to acknowledge A. Pikovsky and M. Zaks for useful discussions.
This work has been financially supported by the EU project COSMOS (642563).

\section*{References}
\bibliography{biharm}
\bibliographystyle{iopart-num}

\end{document}